# Muon Acceleration Concepts for NuMAX: "Dual-use" Linac and "Dogbone" RLA

**S. Alex Bogacz**

*Jefferson Lab,*
*12050 Jefferson Avenue, Newport News, VA 23606, USA*
*E-mail*: bogacz@jlab.org

ABSTRACT: We summarize the current state of a concept for muon acceleration aimed at a future Neutrino Factory. The main thrust of these studies was to reduce the overall cost while maintaining performance by exploring the interplay between the complexity of the cooling systems and the acceptance of the accelerator complex. To ensure adequate survival for the short-lived muons, acceleration must occur at high average gradient. The need for large transverse and longitudinal acceptances drives the design of the acceleration system to an initially low RF frequency, e.g., 325 MHz, which is then increased to 650 MHz as the transverse size shrinks with increasing energy. High-gradient normal conducting RF cavities at these frequencies require extremely high peak-power RF sources. Hence superconducting RF (SRF) cavities are chosen. We consider two cost effective schemes for accelerating muon beams for a stageable Neutrino Factory: exploration of the so-called "dual-use" linac concept, where the same linac structure is used for acceleration of both $H^-$ and muons and, alternatively, an SRF-efficient design based on a multi-pass (4.5) "dogbone" RLA, extendable to multi-pass FFAG-like arcs.

# Contents

## 1. Introduction

We present key concepts for muon acceleration aimed at a future Neutrino Factory. We follow the leading design options recently identified for NuMAX [1] (5 GeV Neutrino Factory). The main thrust of these studies was to reduce the overall cost while maintaining performance through exploring the interplay between the complexity of the cooling systems and the acceptance of the accelerator complex.

To ensure adequate survival for the short-lived muons, acceleration must occur at high average gradient. The accelerator must also accommodate the phase-space volume occupied by the beam after the cooling channel, which is still large. The need for large transverse and longitudinal acceptances drives the design of the acceleration system to an initially low RF frequency, e.g., 325 MHz, which is then increased to 650 MHz as the transverse size shrinks with increasing energy. High-gradient normal conducting RF cavities at these frequencies require extremely high peak-power RF sources. Hence superconducting RF (SRF) cavities are preferred. In the following we choose an SRF gradient of 20 MV/m at 325 MHz and subsequently 25 MV/m at 650 MHz, which allows survival of about 85% of the muons as they are accelerated to 5 GeV. Significant groundwork, schemes and fundamental building blocks were already laid by the IDS-NF [2] efforts and further refined by MASS [1].

We consider two cost effective schemes for accelerating muon beams for a stageable 5 GeV Neutrino Factory:

- Scheme I: SRF efficient design based on a multi-pass (4.5) "Dogbone" RLA, extendable to multi-pass FFAG-like arcs.

- Scheme II: Exploration of the so-called "dual-use" linac concept, where the same linac structure is used for acceleration of both $H^-$ and muons.

Both design options are illustrated schematically in Fig. 1.

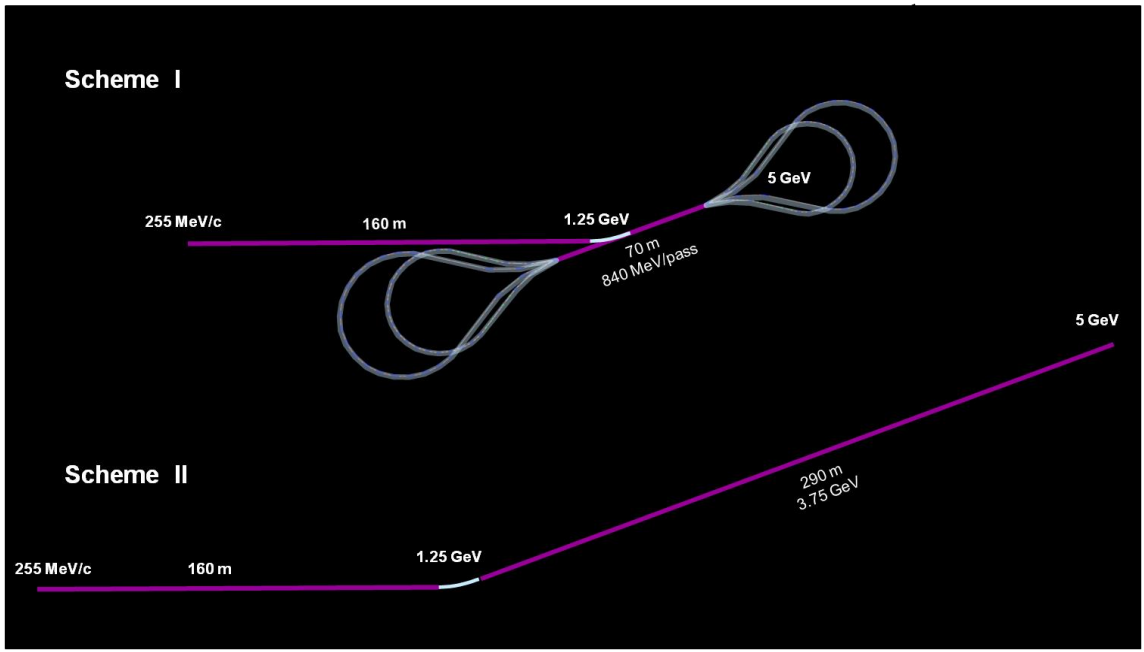

Figure 1. Two design options for NuMAX: schematic views of the overall accelerator complex.

## 2. Muon Accelerator Complex: Overview

The proposed muon accelerator complex consists of a single-pass, superconducting linac with 325 MHz RF cavities, accelerating muons to 1.25 GeV, that captures the large muon phase-space coming from the bunch rotator (in the case of NuMAX), or the factor-of-2 smaller phase-space after the cooling channel (in the case of the full-luminosity Neutrino Factory). The large acceptance of the linac requires large apertures and tight focusing. This, combined with moderate beam energies, favors solenoid rather than quadrupole focusing for the entire linac [3]. The initial single-pass linac accelerates muons to sufficiently relativistic energy, 1.25 GeV, beyond which acceleration using a more efficient and compact higher frequency, 650 MHz, linac structure becomes feasible.

For simultaneous acceleration of both muon charge species, the transition to the 650 MHz linac requires a half-wavelength path-length delay for one of the muon species. This is facilitated by a path-length delay double chicane, in which muons of different charges follow alternative chicane legs, different in length by a half-wavelength at 650 MHz, thus 230.6 mm.

Then for further acceleration (from 1.25 GeV to 5 GeV), our design study branches into two alternative paths: Scheme I, in which the linac is followed by a 4.5-pass, recirculating linear accelerator (RLA) in a "dogbone" configuration and Scheme II, in which a single-pass linac shared between muons and H⁻ is used—the so called "dual-use" linac. In both schemes, acceleration beyond 1.25 GeV continues using a more compact and efficient 650 MHz SRF structure, while adiabatically decreasing the phase-space volume. The two alternative concepts are described in detail in the following sections.

## 3. Initial Single-Pass Linac

A single-pass linac starting at 255 MeV/c (276 MeV total energy) raises the total energy of the muons to 1.25 GeV. This makes the muons sufficiently relativistic to facilitate further acceleration in the RLA or in the dual-use linac. The initial phase-space of the beam, as delivered by the muon front-end, is characterized by a significantly large energy spread; the linac has been designed so that it first confines the muon bunches in longitudinal phase-space, then adiabatically superimposes acceleration over the confinement motion, and finally boosts the confined bunches to 1.25 GeV. To achieve a manageable beam-size in the front-end of the linac, short focusing cells (with one 2-cell cavity) are used for the first 22 cryomodules. The beam size is adiabatically damped with acceleration, allowing the short cryomodules to be replaced with 30 intermediate length cryomodules (with one 4-cell cavity). Consequently, the linac is split into two consecutive sections (referred to as the short- and middle-cell linac sections), each section being built of a particular type of cryomodule as shown in Figure 2 [4]. Each linac section is configured with periodic FOFO cells, matched at the section junctions, as illustrated in Figure 2. Periodicity within each section is maintained by scaling the solenoid fields in consecutive cryomodules linearly with increasing momentum [5]. The cavity iris radius limits the physical aperture of the linac. The radius of 15 cm, matched with a 2.5 sigma beam envelope, defines the transverse acceptance (normalized) of the linac as 20 mm·rad, as shown in Figure 2 (bottom).

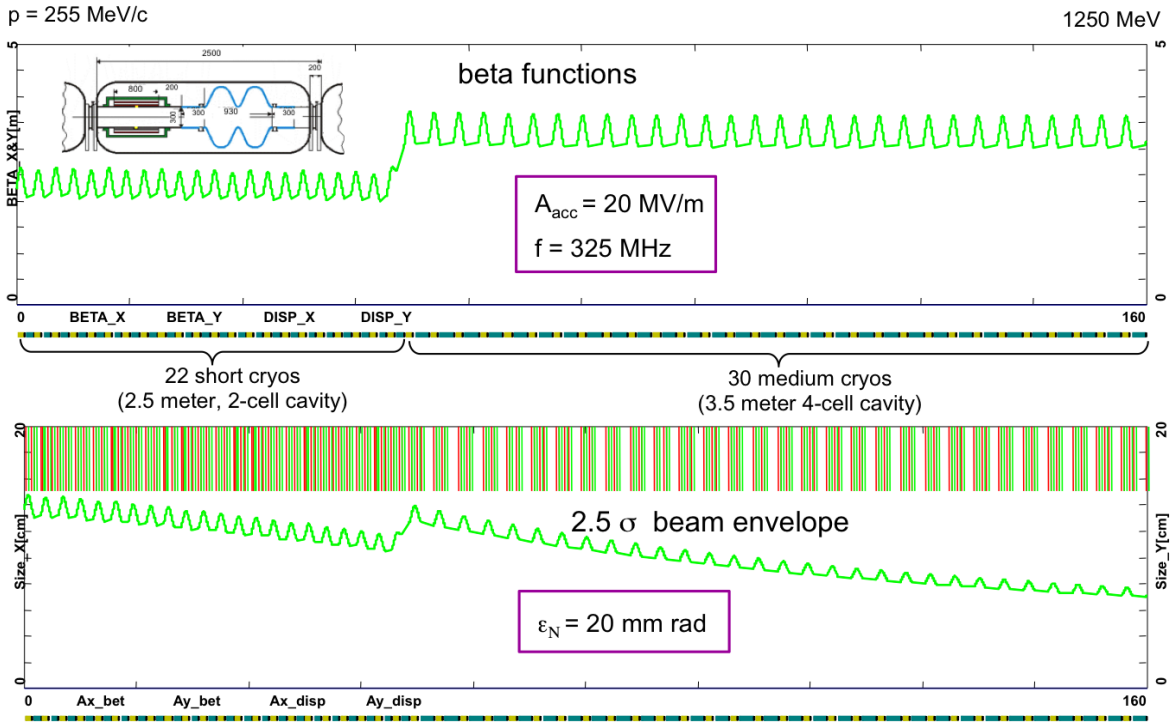

Figure 2. Transverse FOFO optics (top) and the beam envelope of the initial linac; the short- and middle-cell periodic sections are uniformly matched across the junction. The physical aperture radius of 15 cm defines the transverse acceptance (normalized, rms) as 20 mm·rad (bottom).

One of the main requirements of the initial single-pass linac is to compress adiabatically the longitudinal phase-space volume in the course of acceleration. The initial longitudinal acceptance of the linac (chosen to be 2.5 σ) calls for "full bucket" acceleration, with an initial momentum acceptance $\Delta p/p = \pm 28\%$ and bunch length $\Delta\varphi = \pm 97.5°$ (in RF phase). To perform adiabatic bunching one needs to drive rather strong synchrotron motion along the linac [6]. The profile of the RF-cavity phases is organized so that the phase of the first cavity is shifted by 84° (off crest) and then the cavity phase is gradually changed to about 10° (off crest) by the end of the linac, as shown in Figure 3a.

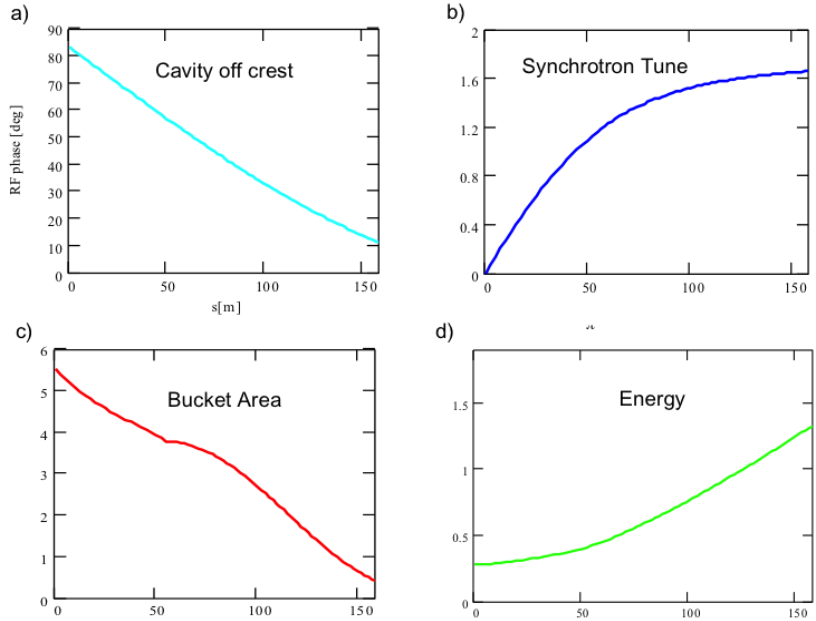

Figure 3. Longitudinal matching of the initial single-pass linac: a) cavity phasing starts far off-crest to capture the beam at low energy, then it moves closer to the crest as the longitudinal acceptance improves with increasing energy; b) the synchrotron phase advances by over one and a half periods from the beginning to the end of the linac; c) the bucket area is initially very large to capture a maximum longitudinal acceptance of 150 mm; it decreases as the bunch is compressed moving closer to crest; d) the rate of energy gain increases as the bunch moves closer to crest.

In the initial part of the linac, when the beam is still not relativistic, the far off-crest acceleration induces rapid synchrotron motion (one and a half full periods along the linac, as shown in Figure 3b), which allows bunch "head" and "tail" to switch places within the RF bucket three times during the course of acceleration. This process [6] is essential for averaging energy spread within the bunch, which ultimately yields desired bunch compression in both bunch-length and momentum spread, as illustrated in Figure 4. To maximize the longitudinal acceptance, the initial position of the bunch is shifted relative to the center of the bucket, to keep the beam boundary inside the separatrix, as illustrated in Figure 4a. The synchrotron motion also suppresses the sag in acceleration [4] for the bunch head and tail as seen in Figure 4b.

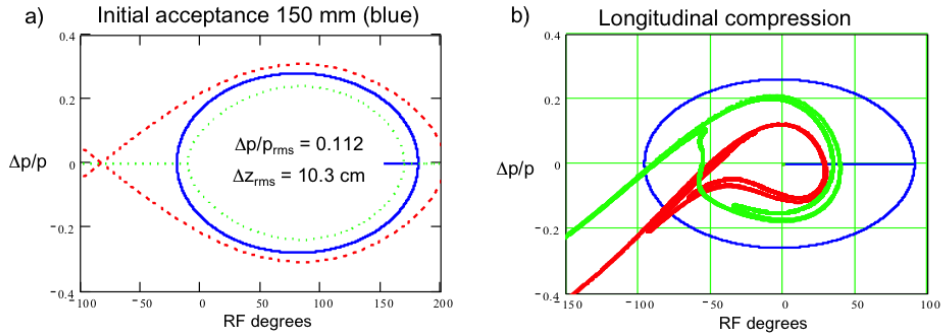

Figure 4. a) Longitudinal acceptance of 150 mm matched inside the separatrix and optimized for "full bucket" acceleration; b) longitudinal compression in the course of acceleration: initial acceptance (blue) propagated through the first part of the linac (green) and finally fully compressed bunch at the end of the linac (red).

## 4. Delay/Compression Double Chicane

To maintain simultaneous acceleration of both muon charge species, the transition from the 325 MHz to the 650 MHz linac requires a half-wavelength path-length delay for one of the muon species in order to re-establish synchronous acceleration at the new RF frequency, as depicted schematically in Figure 5a. Furthermore, transition to higher RF frequency (650 MHz) introduces an abrupt change in the bucket shape; the new bucket is about a factor of 2 shorter and slightly higher, since larger RF gradient is achievable (25 MV/m at 650 MHz vs. 20 MV/m at 325 MHz), as illustrated in Figure 5b.

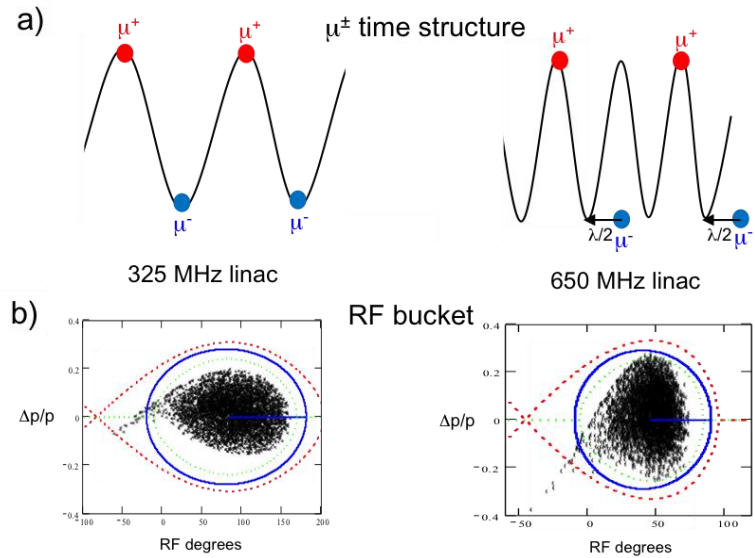

Figure 5. a) Time structure of muon bunches; doubling the RF frequency requires a path-length delay of $\lambda/2$ for $\mu^-$ to put them into accelerating buckets of the new frequency; b) change of the bucket length (a factor of 2 shorter when the RF frequency is doubled) requires compression of the bunch-length by about factor of 2 in order to fit the bunch into the new bucket.

Both effects are addressed by a dedicated path-length delay/compression double chicane, configured with six 30° bends, where muons of different charges follow alternative chicane legs, different in length by the half-wavelength at 650 MHz, as illustrated in Figure 6.

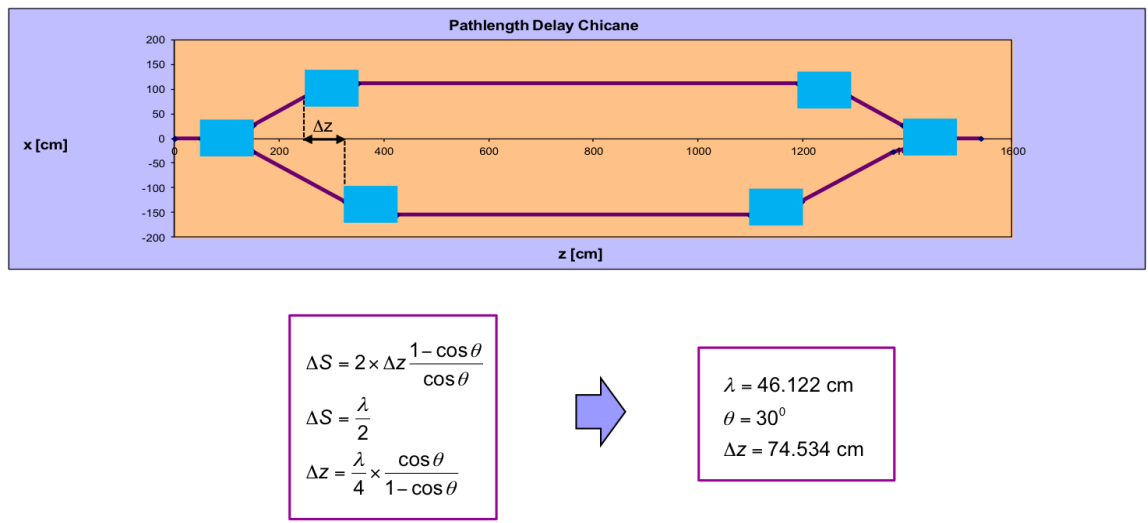

Figure 6. Layout of the path-length delay double chicane, in which muons of different charges follow alternative chicane legs, different in length by a half-wavelength at 650 MHz.

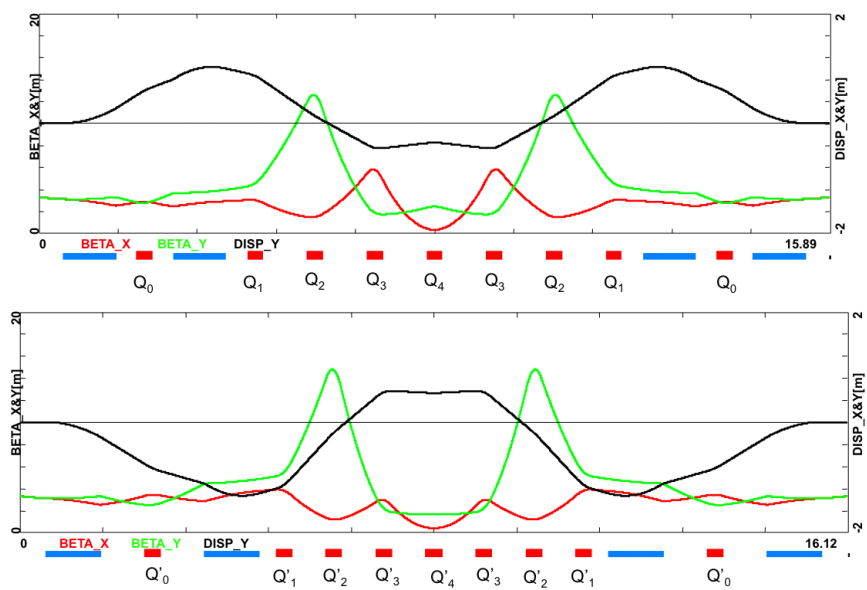

Figure 7. Optics for the two "legs" of the double chicane (top/bottom), featuring mirror-symmetric vertical achromats, with significant momentum compaction, $M_{56} = -1$ m, matched to the same Twiss functions at the ends.

Each leg of the chicane is configured with 9 quadrupoles powered in mirror-symmetric fashion to provide 5 independent parameters required for tuning of the two betas and two alphas and to suppress the vertical dispersion (2 + 2 + 1), as illustrated in Figure 7. Furthermore, this style of optics features large negative momentum compaction, which, combined with an energy chirp in the upstream linac, works as an effective quarter-wave rotator in longitudinal phase-space, yielding the bunch-length reduction required for acceleration in the 650 MHz linac, downstream of the chicane. The dynamics of the above process was studied numerically using a particle tracking simulation via the matrix-based code OptiM [9], as illustrated in Figure 8. In the simulation we have assumed a particle distribution that is Gaussian in 6D phase space with the tails of the distribution truncated at 2.5 sigma, which corresponds to the beam acceptance. Now the muon beam is ready for further acceleration in a 650 MHz linac structure; either in a single-pass "dual-use" linac (Scheme II), or in a multi-pass "dogbone" RLA (Scheme I). Both schemes are described in detail in the following sections.

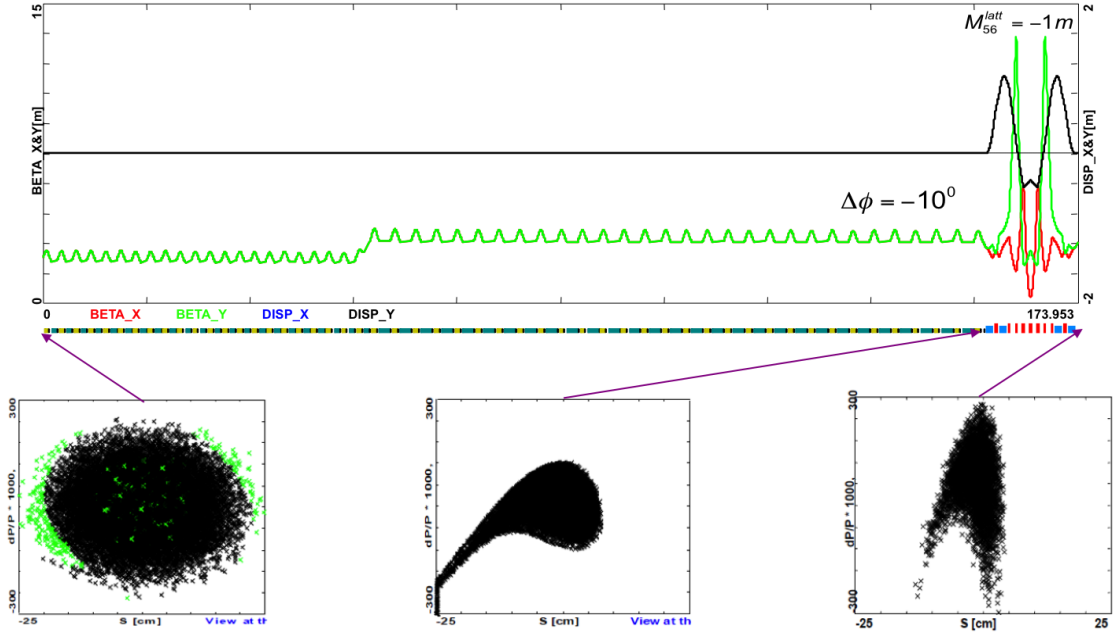

Figure 8. Snapshot of the initial longitudinal phase-space, corresponding to full longitudinal acceptance of 150 mm (left), transported through the single-pass linac, generating energy chirp by being phased off-crest (middle snapshot), and finally compressed bunch (factor of two shorter) through energy correlated momentum compaction across the chicane (right snapshot).

## 5. "Dual-use" Linac

The main virtue of the "dual-use" linac is to re-use a large SRF linac structure, which is already needed for a proton driver (used for pion and ultimately muon production), to accelerate muons from 1.25 GeV to 5 GeV. This would offer significant cost savings, if muon and proton acceleration are compatible, which will be addressed at the end of this section.

Following the double chicane, at 1.25 GeV, the transverse beam size is small enough that the beam may be accelerated using a more compact and efficient 650 MHz SRF structure with aperture radius (cavity iris) of 7.5 cm. The initial beam conditions at 1.25 GeV still favor solenoid rather than quadrupole focusing for the initial part of this linac. The front-end is therefore configured with 28, 4-meter-long cryomodules, each containing two 4-cell cavities and a superconducting counter-wound solenoid (similar in layout to the previously described 325 MHz cryomodules). The beam size is further damped with acceleration, such that at around 2.5 GeV a quadrupole based FODO structure becomes more efficient than the solenoid FOFO optics. Consequently, the linac continues with 28, 4-meter-long, FODO cells, containing two 4-cell cavities and two quads. Then, at about 3.8 GeV, the transverse beam size becomes small enough that one may switch to longer and more efficient FODO cells (6-meter-long) with four 4-cell cavities. Eleven such cells complete the 5 GeV linac. The linac sections, configured with periodic either FOFO or FODO cells, are uniformly matched at the junctions, as illustrated in Figure 9.

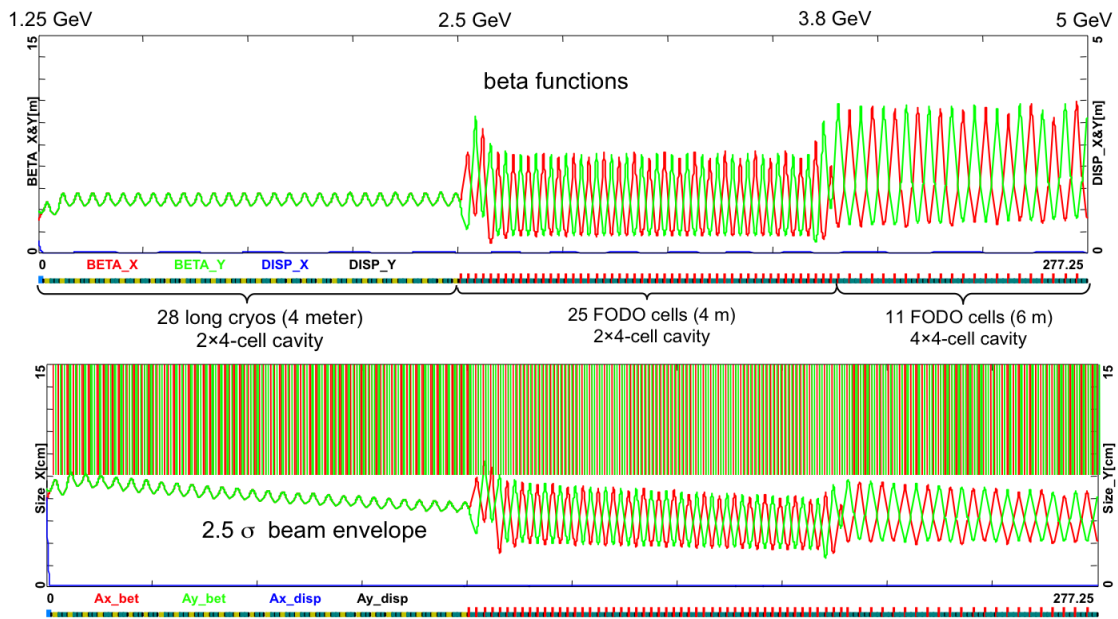

Figure 9. "Dual-use" linac based on 650 MHz SRF: initial linac section with solenoid based FOFO optics, followed by two FODO-style sections with quadrupole focusing (top), and the beam envelope of the "dual-use" linac: three styles of periodic sections are uniformly matched across the junctions. The physical aperture radius of 7.5 cm is compatible with the transverse acceptance (normalized) of the initial single-pass linac, 20 mm·rad (bottom).

Ultimately, one needs to look into the above "dual-use" linac design from the H⁻ side. Comparison of the Twiss functions for both muons and protons at appropriate energies is illustrated in Figure 10. It shows that acceleration of both species in the same linac structure is quite compatible, with less then a factor of three increase in beta functions for H⁻.

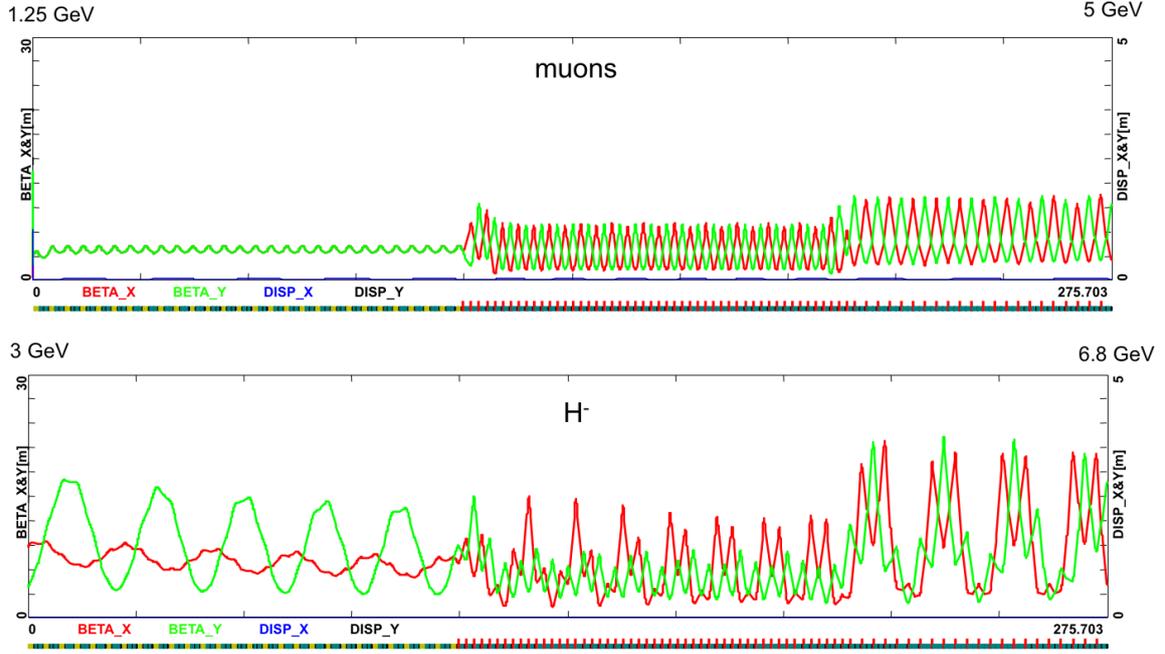

Figure 10. Compatibility of muon and H$^-$ acceleration in the "dual-use" linac: beta functions for muons accelerated from 1.25 GeV to 5 GeV (top) and the corresponding Twiss functions for H$^-$ beam accelerated from 3 GeV to 6.8 GeV (bottom).

## 6. "Dogbone" RLA

The main virtue of the multi-pass RLA option is its very efficient use of an expensive SRF linac. The "dogbone" RLA is designed to accelerate simultaneously the µ$^+$ and µ$^-$ beams from 1.25 GeV to 5 GeV to further compress and shape the longitudinal and transverse phase-space [6]. The beam is injected from the single-pass linac via the double chicane (as discussed in Section 3). The injection point into the "dogbone" RLA coincides with the middle of the multi-pass linac to minimize the effect of phase slippage for the initially 1.25 GeV muon beam accelerated in linacs that are phased for speed-of-light particles. At the ends of the RLA linacs the beams need to be directed into the appropriate energy-dependent (pass-dependent) "droplet" arc for recirculation [7]. The above configuration has already been introduced in Figure 1 (top). The 650 MHz SRF linac is configured with smaller aperture (7.5 cm radius), higher gradient (25 MV/m) cavities.

### 6.1 Multi-pass Linac

The injection energy into the RLA and the energy gain per RLA linac (840 MeV) were chosen so that a tolerable level of RF phase slippage along the linac could be maintained (~ 20° in RF phase). To suppress chromatic effects, 90° FODO optics is used as a building block for both the linac and the return arcs [5]. In the linac, the 4-meter-long FODO cells are configured with 4-cell SRF cavities between each of the quadrupoles, except that there are no cavities between the central D magnet and its adjacent F quadrupoles. The layout and optics of the linac periodic FODO structure are shown in Figure 11a. The injection chicane magnets are located at the middle of the linac (marked in blue).

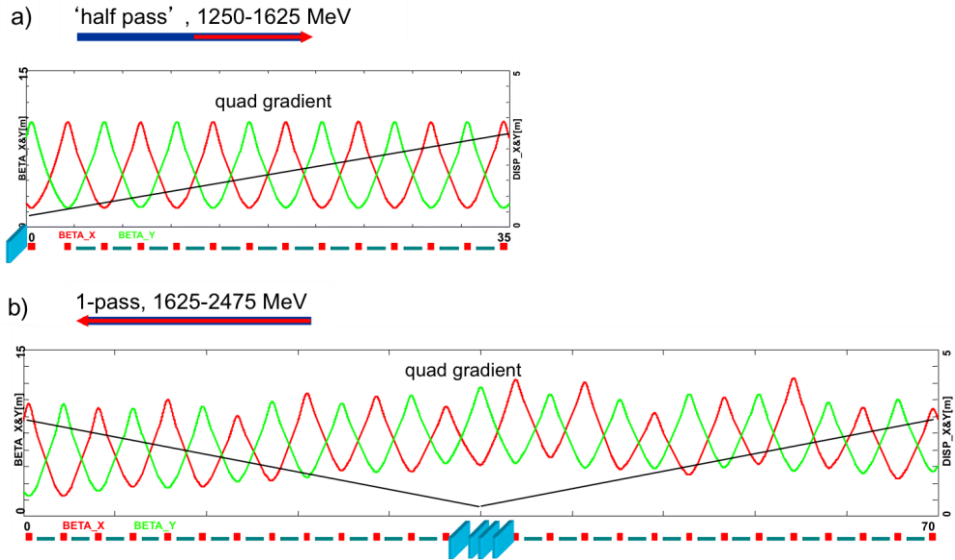

Figure 11. "Bisected" linac optics: the quadrupole gradients are scaled up with momentum for the first half of the linac, then they are mirror reflected in the second half: (a) periodic FODO structure set for the lowest energy "half-pass" through the linac; (b) linac optics for the first "full pass"; underfocusing effects in the first half of the linac are mitigated by reversing the focusing profile in the second half.

The focusing profile along the linac was chosen so that beams with a large energy spread could be transported within the given aperture. Since the beam is traversing the linac in both directions, a "bisected" focusing profile was chosen for the multi-pass linac [7]. Here, the quadrupole gradients scale up with momentum to maintain 90° phase advance per cell for the first half of the linac (see Figure 11a), then are mirror reflected in the second half, as illustrated in Figure 11b. Performance of the bisected linac optics for all 4.5 passes is illustrated in Figure 12.

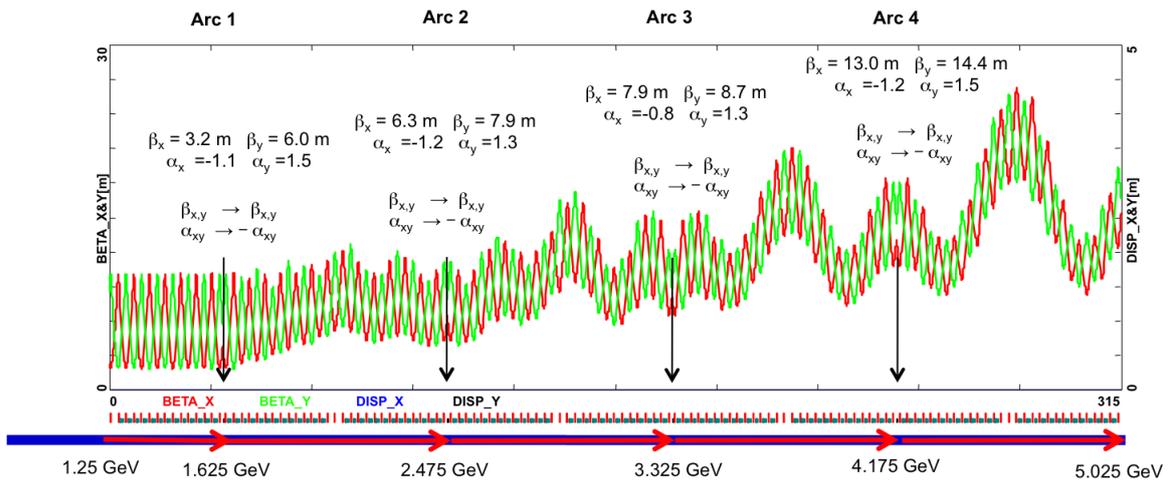

Figure 12. "Bisected" linac optics for all passes, with mirror symmetric arcs inserted as point matrices (arrows). The virtue of the optics is the appearance of distinct nodes in the beta beat-wave at the ends of each pass (where the arcs begin), which limits the growth of initial betas at the beginning of each subsequent droplet arc (Arc 1–4), hence eases linac-to-arc matching.

## 6.2 "Droplet" Arc

At the ends of the RLA linac, the beams need to be directed into the appropriate energy-dependent (pass-dependent) droplet arc for recirculation. The entire droplet-arc architecture [8] is based on 90° phase-advance cells with periodic beta functions, as depicted in Figure 13. For practical reasons, horizontal rather than vertical beam separation has been chosen. Rather than suppressing the horizontal dispersion created by the spreader, it has been matched to that of the outward arc. This is partially accomplished by removing one dipole (the one furthest from the spreader) from each of the two cells following the spreader. To switch from outward to inward bending, three transition cells are used, from which the four central dipoles are removed. The two remaining dipoles at the ends bend in the same direction as the dipoles to which they are closest. The transition region, across which the horizontal dispersion switches sign, is therefore composed of two such cells. To facilitate simultaneous acceleration of both $\mu^+$ and $\mu^-$ bunches; mirror symmetry is imposed on the droplet arc optics (oppositely charged bunches move in opposite directions through the arcs). This puts a constraint on the exit/entrance Twiss functions for two consecutive linac passes, namely: $\beta_{n\,out} = \beta_{n+1\,in}$ and $\alpha_{n\,out} = -\alpha_{n+1\,in}$, where n = 0, 1, 2... is the pass index. The complete droplet arc optics for the lowest-energy pair of arcs is shown in Figure 13.

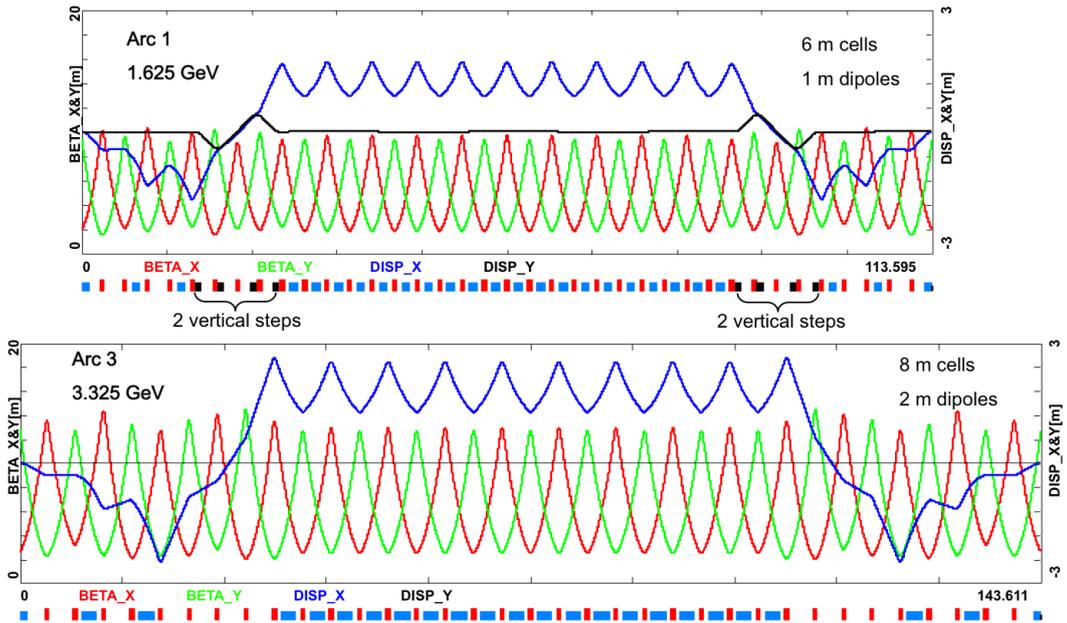

Figure 13. Droplet arc optics for a pair of arcs on one side of the "dogbone," Arc 1 and Arc 3.

All higher arcs are based on the same principle as Arc 1, with gradually increasing cell length (and dipole magnet length) to match naturally to the increasing beta functions dictated by the multi-pass linac. The quadrupole strengths in the higher arcs are scaled up linearly with momentum to preserve the 90° FODO lattice. The physical layout of the above pair of droplet arcs is illustrated in Figure 14. The momentum compaction is relatively large (6.5 m), which guarantees significant rotation in longitudinal phase-space as the beam passes through the arc. This effect, combined with off-crest acceleration in the subsequent linac, yields further compression of the longitudinal phase-space as the beam is accelerated.

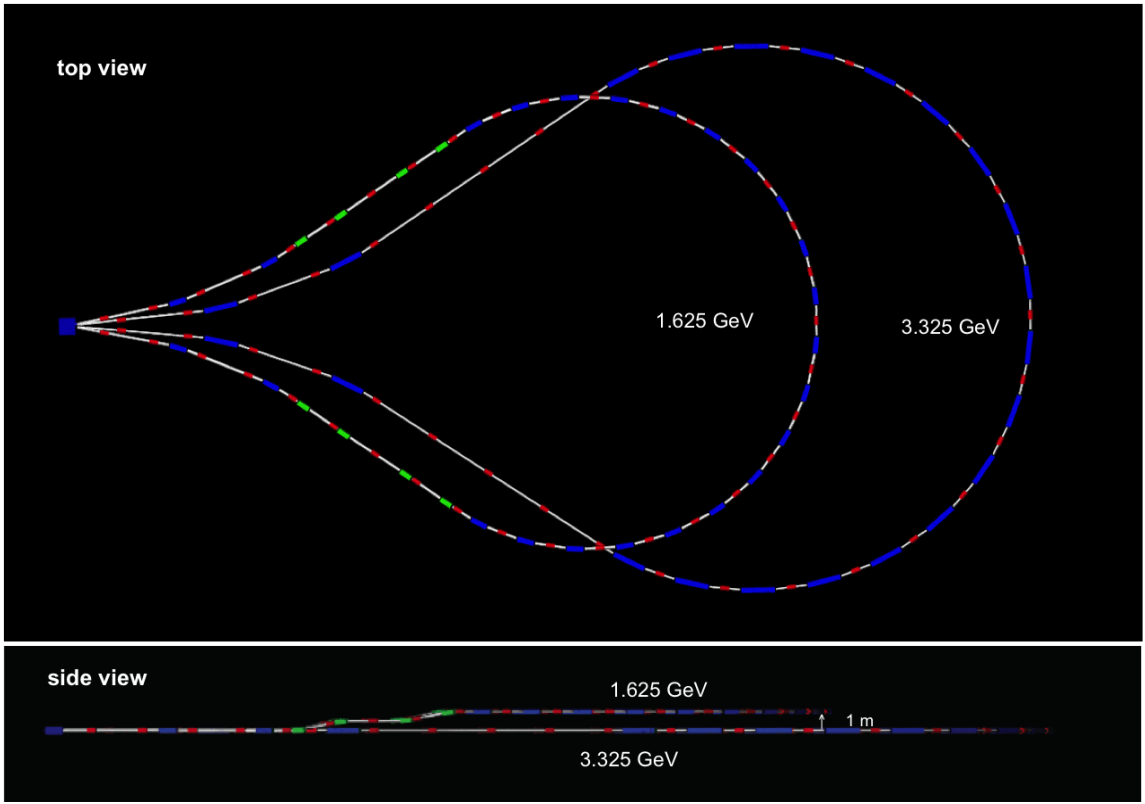

Figure 14. Layout of a pair of arcs on one side of the "dogbone" RLA: Arc 1 and Arc 3, top and side views, showing vertical two-step "lift'" of the middle part of Arc 1 to avoid interference with Arc 3.

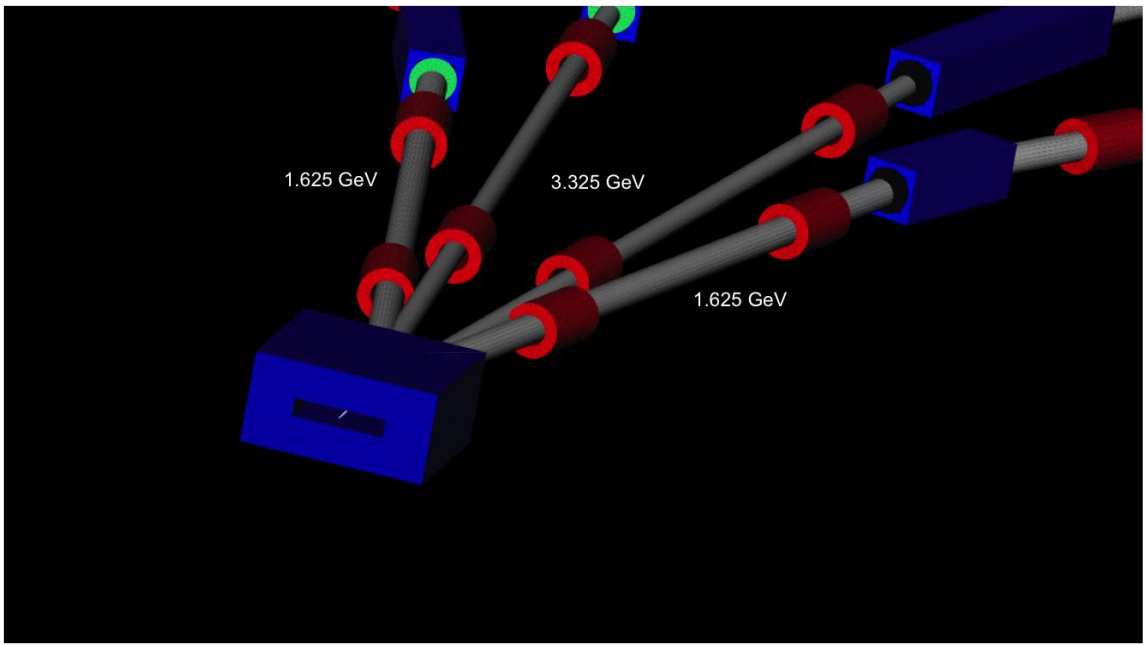

Figure 15. Switchyard layout, featuring a common dipole (blue), separating beams horizontally and directing them to the corresponding arcs. Initial quadrupoles (red) define Spreader/Recombiner optics.

**6.3 Outlook**

In summary, a "dogbone" RLA is a fast, compact and efficient way of accelerating muon beams to medium and high energies by reusing the same linac for multiple passes, where the different energy passes coming out of the linac are separated and directed into individual (single energy) return arcs for recirculation. However, each pass through the linac requires a separate fixed energy arc, hence increasing the complexity, size, and cost of the RLA.

As an alternative, we have also recently proposed a novel return arc optics design based on linear combined function magnets with variable dipole and quadrupole field components. This allows two consecutive passes with very different energies to be transported through the same string of magnets [10-13]. A proof-of-principle droplet arc composed of identical 60° bending cells—two outward- and five inward-bending cells—has been presented [11]. A basic building block of this proposed optics scheme is illustrated in Figure 16, which shows solutions for the periodic orbit, dispersion, and beta functions of the outward-bending supercell at 1.2 and 2.4 GeV/c, respectively.

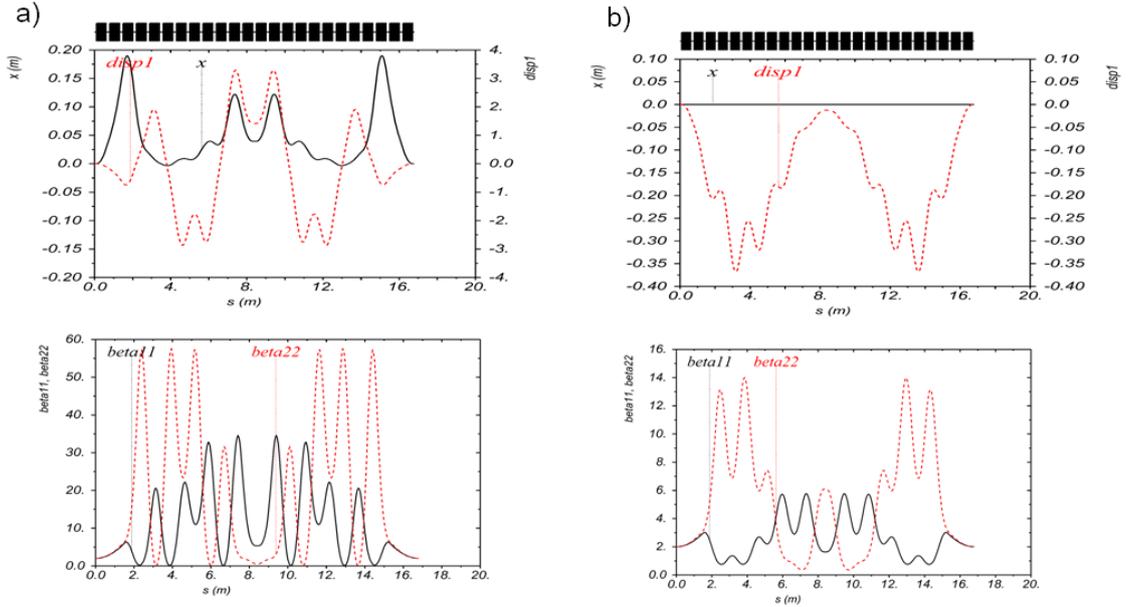

Figure 16. Periodic orbits, dispersion (top) and beta functions (bottom) of a) 1.2 GeV/c and b) 2.4 GeV/c outward bending supercells.

Such a solution [14] combines compactness of design with all the advantages of a linear NS-FFAG [13], namely, large dynamic aperture and momentum acceptance essential for large-emittance muon beams, no need for complicated compensation of non-linear effects, and a simpler combined-function magnet design with only dipole and quadrupole field components. The scheme utilizes only fixed magnetic fields, including those for injection and extraction.

We are currently studying the dynamic aperture and momentum acceptance of the arc. Earlier studies with a similar linear lattice yielded promising results [14]. We will investigate chromatic effects and, if necessary, implement their control and compensation. Another important aspect that requires investigation is the design sensitivity to magnet misalignments and magnetic field errors. Establishing tolerance levels on these errors is crucial for the costing of large aperture magnets.

## 7. Summary and Closing Remarks

The design study we have presented explored two alternative paths: Scheme I, in which the linac is followed by a 4.5-pass, recirculating linear accelerator (RLA) in a "dogbone" configuration, and Scheme II, in which a single pass linac shared between muons and H⁻ is used—the so called "dual-use" linac.

One needs to emphasize that the final choice for NuMAX, favoring the dual-use linac, was based solely on practical considerations such as cost saving and operability. The significant cost saving aspect of the "dual-use" linac is rather obvious, since a large SRF linac structure, already needed for a proton driver (pion and ultimately muon production), is re-used to accelerate muons from 1.25 GeV to 5 GeV. That overshadowed the elegance and flexibility of the RLA scheme.

The key advantage of a multi-pass RLA is its very efficient use of an expensive SRF linac, making the scheme elegant and flexible. Furthermore, the "dogbone" RLA is well suited for simultaneous acceleration of the $\mu^+$ and $\mu^-$ beams, while further compressing and shaping the longitudinal and transverse phase-space. Its inherently large momentum compaction, combined with off-crest acceleration in the subsequent pass through the linac, facilitates significant rotation in the longitudinal phase-space, which yields further compression of the longitudinal phase-space as the beam is accelerated. This feature makes it much superior to a straight linac. Finally, the "dogbone" scheme can be extended into a cascade of RLAs, as was envisioned within MASS [1]: the 5 GeV RLA (NuMAX) was followed first by a larger RLA to 63 GeV (Higgs Factory), and finally by another RLA to reach TeV-scale energies (multi-TeV Muon Collider). That is why a multi-pass (>8) "dogbone" RLA, configured with the proposed FFAG-like arcs, would be the technology of choice for the extreme muon acceleration needed for future muon facilities (TeV scale and beyond).